 \documentclass[sigconf]{acmart}

\makeatletter
\def\@affiliationfont{\normalsize\normalfont}
\makeatother

\copyrightyear{2026}
\acmYear{2026}
\setcopyright{cc}
\setcctype{by}
\acmConference[WWW '26] {Proceedings of the ACM Web Conference 2026}{April 13--17, 2026}{Dubai, United Arab Emirates.}
\acmBooktitle{Proceedings of the ACM Web Conference 2026 (WWW '26), April 13--17, 2026, Dubai, United Arab Emirates}
\acmISBN{979-8-4007-2307-0/2026/04}
\acmDOI{10.1145/3774904.3792156}
\usepackage{amsmath,amsfonts}
\usepackage{bbm}
\usepackage{mathtools}
\usepackage{enumitem}
\usepackage{multirow}
\usepackage{tabularx,booktabs}
\usepackage[table,dvipsnames]{xcolor}
\definecolor{RowBlue}{HTML}{E6F2FF}
\newcolumntype{Y}{>{\centering\arraybackslash}X} % 可伸缩的居中列

\newcommand{\tbf}[1]{\textbf{#1}}
\newcommand{\udl}[1]{\underline{#1}}

\begin{document}

\title{DiffGRM: Diffusion-based Generative Recommendation Model}

% 贡献与通讯作者标记
\newcommand{\cofirst}{\authornote{Equal contribution.}}
\newcommand{\cofirstmark}{\authornotemark[1]}

\newcommand{\corr}{\authornote{Corresponding author.}}
\newcommand{\corrmark}{\authornotemark[2]} % 对应作者现在是第2个脚注（†）

% 实习脚注：只打第3个脚注标记（‡），不在这里生成脚注文本
\newcommand{\internmark}{\authornotemark[3]}

% 只插入作者脚注文本（不生成角标），用于补上第3条脚注内容
\makeatletter
\newcommand{\authornotetextonly}[1]{%
  \if@ACM@anonymous\else
    \g@addto@macro\@authornotes{%
      \stepcounter{footnote}\footnotetext{#1}}%
  \fi}
\makeatother

% 机构在第一行；城市+国家在第二行（同一行）
\newcommand{\ksaffil}{%
  \affiliation{%
    \institution{Kuaishou Technology}%
    \city{Beijing}%
    \country{China}%
  }%
}

\newcommand{\unaaffil}{%
  \affiliation{%
    \institution{Independent}%
    \city{Beijing}%
    \country{China}%
  }%
}

\author{Zhao Liu}\cofirst
\ksaffil
\email{liuzhao09@kuaishou.com}

\author{Yichen Zhu}\cofirstmark\internmark
\ksaffil
\email{zhuyichen03@kuaishou.com}

\author{Yiqing Yang}\cofirstmark
\ksaffil
\email{yangyiqing06@kuaishou.com}

\author{Xiao Lv}\corrmark
\ksaffil
\email{lvxiao03@kuaishou.com}

\author{Guoping Tang}
\ksaffil
\email{tangguoping@kuaishou.com}

\author{Rui Huang}
\ksaffil
\email{huangrui06@kuaishou.com}

\author{Qiang Luo}\corr
\ksaffil
\email{luoqiang@kuaishou.com}

\author{Ruiming Tang}
\ksaffil
\email{tangruiming@kuaishou.com}

\author{Kun Gai}
\unaaffil
\email{gai.kun@qq.com}

\author{Guorui Zhou}\corrmark
\ksaffil
\email{zhouguorui@kuaishou.com}

\authornotetextonly{Work done during internship at Kuaishou.}

\renewcommand{\shortauthors}{Zhao Liu et al.}

\begin{abstract}
Generative recommendation (GR) is an emerging paradigm that represents each item via a tokenizer as an $n$-digit semantic ID (SID) and predicts the next item by autoregressively generating its SID conditioned on the user's history.
However, two structural properties of SIDs make ARMs ill-suited. First, intra-item consistency: the $n$ digits jointly specify one item, yet the left-to-right causality trains each digit only under its prefix and blocks bidirectional cross-digit evidence, collapsing supervision to a single causal path. Second, inter-digit heterogeneity: digits differ in semantic granularity and predictability, while the uniform next-token objective assigns equal weight to all digits, overtraining easy digits and undertraining hard digits.
To address these two issues, we propose DiffGRM, a diffusion-based GR model that replaces the autoregressive decoder with a masked discrete diffusion model (MDM), thereby enabling bidirectional context and any-order parallel generation of SID digits for recommendation.
Specifically, we tailor DiffGRM in three aspects: (1) tokenization with Parallel Semantic Encoding (PSE) to decouple digits and balance per-digit information; (2) training with On-policy Coherent Noising (OCN) that prioritizes uncertain digits via coherent masking to concentrate supervision on high-value signals; and (3) inference with Confidence-guided Parallel Denoising (CPD) that fills higher-confidence digits first and generates diverse Top-$K$ candidates.
Experiments show consistent gains over strong generative and discriminative recommendation baselines on multiple datasets, improving NDCG@10 by 6.9\%–15.5\%.
Code is available at: https://github.com/liuzhao09/DiffGRM.

% https://github.com/liuzhao09/DiffGRM
% https://anonymous.4open.science/r/DiffGRM-70FD

\end{abstract}


% ---- Keywords ----

\maketitle

\section{Introduction}

\begin{figure}[t]
\centering
    \includegraphics[width=\columnwidth]{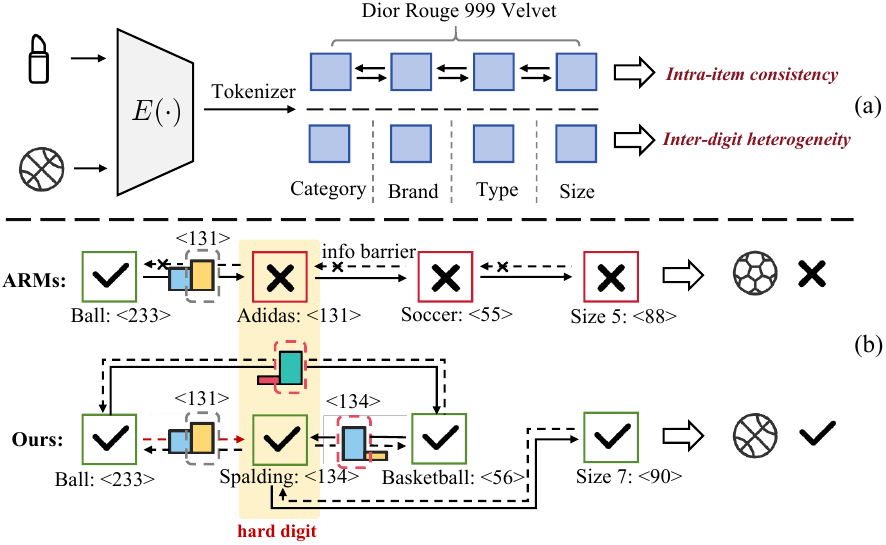}
    \caption{Overview of SID structure and ARM limitations. (a) Intra-item consistency means that the $n$ SID digits jointly specify one item. Inter-digit heterogeneity means that digits encode different semantic granularity and have different prediction difficulty. (b) ARMs decode causally from left to right, which prevents cross-digit verification and can cause errors on hard digits to propagate. Our method updates digits by confidence under bidirectional context.}
    % \caption{Overview of SID structure and ARM limitations. (a) Intra-item consistency means the $n$ SID digits jointly specify one item. Inter-digit heterogeneity means digits carry different semantic granularity and therefore different prediction difficulty. (b) Under ARMs’ causal left-to-right prediction, once the first digit <233> is fixed, the second digit, the hard digit, is evaluated only under this prefix and is mispredicted as <131>. Because the causal constraint forbids using evidence from later digits, subsequent digits are decoded on this incorrect prefix, propagating the error to <55> and <88>. Ours selects digits by confidence and updates in any order. Given <233>, it fills the higher-confidence third digit <56>, then revisits the second digit conditioned on <233> and <56> to correct it to <134>, and finally completes <90>.}
    \label{fig:gen_ppl}
\end{figure}

Generative Recommendation (GR) frameworks have emerged as
a promising direction in recommendation systems ~\cite{DBLP:journals/corr/abs-2404-16924}. 
The core idea is to directly generate the token sequence of target item to complete the recommendation task.
A typical GR framework is composed of two key modules:
(1) Tokenization Module,
which leverages pretrained language models~\cite{t5,bge,gte} to encode item content into dense vectors, then discretizes them into a fixed-length $n$-digit sequence of semantic IDs (SIDs), where each digit is a token selected from a codebook via vector quantization~\cite{DBLP:conf/recsys/SinghV0KSZHHWTC24,TIGER};
and (2) Generation Module,
inspired by GPT-style Transformers~\cite{DBLP:conf/nips/VaswaniSPUJGKP17},
which employs autoregressive models (ARMs) to sequentially predict the token for each digit of the target item's SID.

Unlike typical language modeling, in the GR framework each item is represented by an $n$-digit SID produced by an $n$-layer codebook, with one token assigned to each digit.
This design induces two salient structural properties, as shown in Figure~\ref{fig:gen_ppl}(a). \textbf{Intra-item consistency:} the \(n\) SID digits of an item jointly specify a single item; for example, together they resolve to Dior Rouge 999 Velvet. \textbf{Inter-digit heterogeneity:} the digits encode different semantic granularities, such as Category, Brand, Type, and Size, and therefore differ in predictability and difficulty, yielding easy digits and hard digits\footnote{Hard digits are SID positions with lower conditional predictability given the available context, whereas easy digits have higher predictability. Here, ``digit'' means a position in the SID, and a ``token'' is a codeword assigned to that position.}.

These structural characteristics pose significant challenges for ARMs.
\textbf{First}, the inherent sequential dependency of ARMs creates information barriers and reduces supervision to a single causal path. Causal attention enforces left-to-right dependencies, so each digit is trained only under its prefix context, which makes it hard to exploit intra-item consistency via bidirectional intra-item semantics and cross-digit mutual verification. As a result, the learnable supervision signal is forced to collapse into a single causal path~\cite{DBLP:conf/emnlp/GhazvininejadLL19,xue2025anyordergptmaskeddiffusion}.
\textbf{Second}, prediction difficulty and supervision allocation are imbalanced. Inter-digit heterogeneity means that different SID digits carry unequal semantic load and exhibit very different predictability, yet the next-token objective allocates equal supervision to every digit. This mismatch yields training imbalance: hard digits receive too little signal while easy digits are overtrained.
For example, in Figure~\ref{fig:gen_ppl}(b) the \emph{second} digit (Brand) is the hard digit in this instance, while the \emph{third} digit (Ball Type) is relatively easier, highlighting positional imbalance. Yet the next-token objective gives equal supervision to all digits, undertraining harder digits while overemphasizing easier ones, which hinders accurate SID generation~\cite{DBLP:conf/nips/GoyalLZZCB16,DBLP:conf/acl/ShenCHHWSL16}.

To address these limitations, 
we draw inspiration from the rapid progress in discrete diffusion modeling~\cite{D3PM,DBLP:conf/iclr/GongLF0K23} 
and replace the ARM in GR with a masked diffusion model (MDM). 
MDMs naturally leverage bidirectional context, support parallel generation, 
and provide richer supervision signals through random noising, 
making them better aligned with the structural characteristics of item representations.
Nevertheless, effective use of MDMs in recommendation necessitates task-specific tailoring. 
We therefore introduce the \udl{\tbf{Diff}}usion-based \udl{\tbf{G}}enerative \udl{\tbf{R}}ecommendation \udl{\tbf{M}}odel (DiffGRM), which provides three targeted adaptations across tokenization, training, and inference:

\begin{itemize}[leftmargin=*,nosep]
    \item \textbf{Tokenization level.} 
    Mainstream GR tokenizers often rely on residual quantization (RQ), which induces a strict left-to-right residual dependency between digits. This coupling amplifies positional difficulty imbalance and weakens bidirectional interaction across digits. We adopt \textbf{Parallel Semantic Encoding (PSE)} to decouple digits, enable fully parallel prediction, and balance per-digit information.

    \item \textbf{Training level.}
    Recommendation catalogs are far larger than natural-language vocabularies and highly long-tailed, so most items are rarely observed. With random masking, the available supervision is fragmented, which makes it difficult to learn valid $n$-layer token combinations and increases the risk of invalid SIDs at inference. Beyond sparsity, the supervision space exhibits a combinatorial explosion: the number of masking patterns—and thus target--context signals—grows exponentially with $n$. As quantified in Appendix~\ref{app:sup_details}, an $n$-layer codebook yields $n\,2^{\,n-1}$ signals and requires at least $2^{\,n}-1$ masking configurations for full coverage, which is impractical under realistic budgets. Even with abundant data, exhaustive coverage is infeasible, so the central challenge is to select high-quality supervision signals from this vast candidate space rather than attempting to enumerate it. We therefore propose \textbf{On-policy Coherent Noising (OCN)}, which adds noise to the hardest digits according to the model’s on-policy selection while keeping the unmasked context coherent, thereby focusing the training budget on informative signals and avoiding the exponential scatter of fully random masking.

    \item \textbf{Inference level.}
    In natural language applications, MDMs often rely on greedy decoding to produce a single high-quality response. By contrast, recommendation tasks demand a diverse candidate set of items, so simple greedy decoding is inadequate and calls for decoding strategies tailored to recommendation. To meet this need, we design \textbf{Confidence-guided Parallel Denoising (CPD)}, which fills high-confidence digits first and then completes the remainder via a global parallel beam search, yielding accurate and diverse Top-$K$ SID candidates.

\end{itemize}

Our main contributions are summarized as follows:
\begin{itemize}[leftmargin=*,nosep]
    \item 
    We present DiffGRM, the first diffusion-based generative recommendation framework that replaces the autoregressive decoder with a masked discrete diffusion model over SID digits, removing left-to-right constraints and exploiting bidirectional cross-digit context.
    \item 
    At the training level, we propose OCN to curb the combinatorial explosion of supervision in masked diffusion by coherently masking uncertainty-ranked hard digits, focusing the budget on high-value signals; at the inference level, we introduce CPD, a confidence-guided global parallel beam search that yields diverse Top-$K$ SID candidates.
    \item 
    We achieve state-of-the-art results across multiple public datasets, improving NDCG@10 by 6.9\%–15.5\% over strong generative and discriminative recommendation baselines, demonstrating the accuracy and generalization strength of DiffGRM.
\end{itemize}

\begin{figure*}[t]
  \centering
  \includegraphics[width=\textwidth]{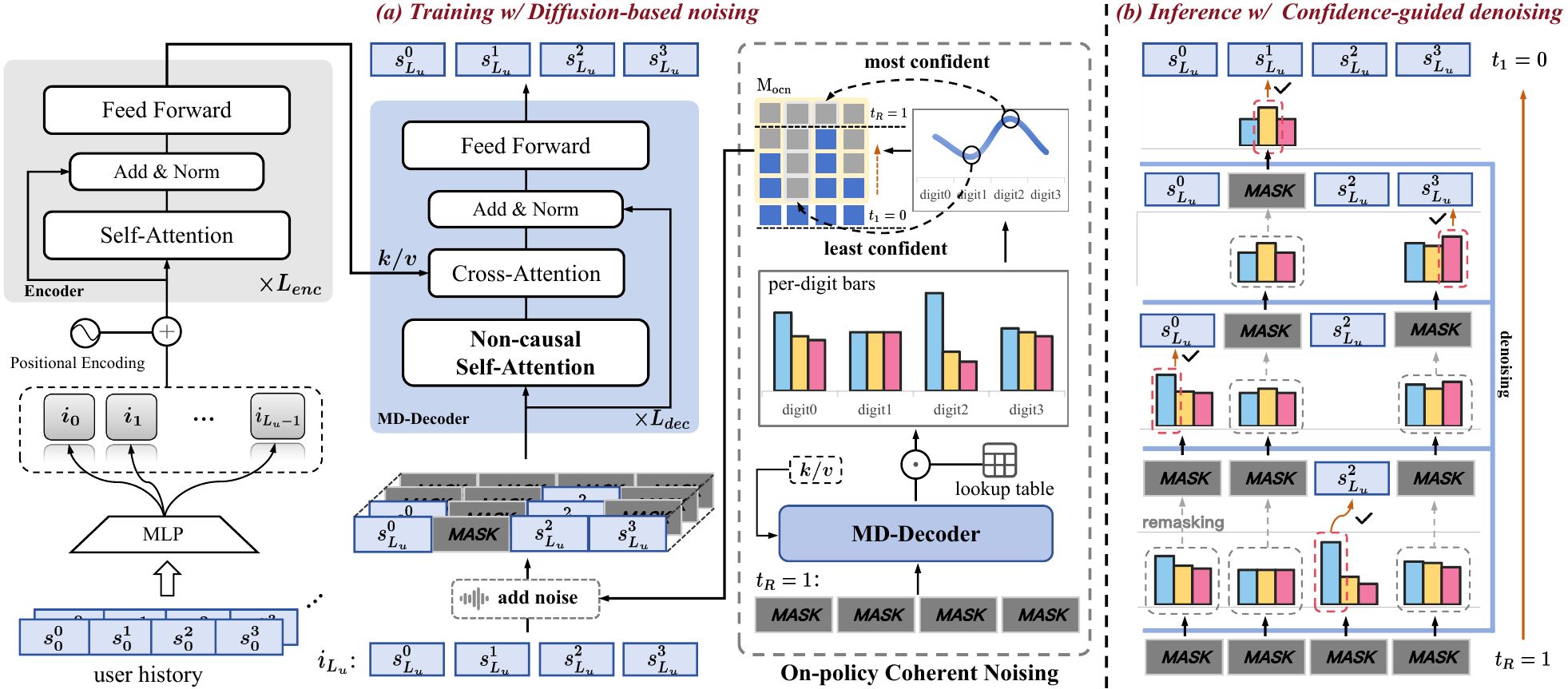}
  \caption{
    Overall framework of DiffGRM. (a) Training: an encoder summarizes the user history, and a masked-diffusion decoder predicts missing digits in parallel under masks scheduled by On-policy Coherent Noising (OCN), which concentrates corruption on harder digits. (b) Inference: Confidence-guided Parallel Denoising (CPD) performs a global parallel beam search from a fully masked input, filling higher-confidence digits first to recover complete SIDs.
    }
  \label{fig:framework}
\end{figure*}

\section{Background}\label{sec:background}

\subsection{SID Generation for Recommendation}\label{subsec:problem}
Given a user's interaction history, we predict the next item by generating its $n$-digit SID. Let $\mathcal{U}$ be the user set and $\mathcal{I}$ the item set. For $u\in\mathcal{U}$, the history is $S_u=[i_1^u,\dots,i_{L_u}^u]$. Each item $i\in\mathcal{I}$ has content features $\mathbf{f}_i$. A semantic encoder $E(\cdot)$ maps them to an embedding $\mathbf{h}_i=E(\mathbf{f}_i)\in\mathbb{R}^d$. We discretize $\mathbf{h}_i$ into an $n$-digit $\mathrm{SID}_i=[s_i^0,\dots,s_i^{n-1}]$ with per-digit codebooks of size $M$, so $s_i^k\in\{0,\dots,M-1\}$. The user history in SID space is $X_u=[\mathrm{SID}_{i_1^u},\dots,\mathrm{SID}_{i_{L_u}^u}]$. Let the target SID be $\mathbf{y}^{*}=\mathrm{SID}_{i_{L_u+1}^u}$. We learn a conditional generator $p_{\theta}(\mathbf{y}\mid X_u)$ and optimize the conditional log-likelihood as:
\begin{equation}
\max_{\theta}\; \mathbb{E}_{u\in\mathcal{U}}\big[\log p_{\theta}(\mathbf{y}^{*}\mid X_u)\big],
\end{equation}
where $p_{\theta}$ models the joint distribution over the $n$ digits given $X_u$, $\mathbf{y}^{*}$ is the SID of the next item for user $u$, and the expectation is over users in $\mathcal{U}$. In our system $p_{\theta}$ is instantiated by a masked-diffusion decoder trained with digit-wise supervision on masked digits.

\subsection{Masked Diffusion for Parallel Token Prediction}\label{subsec:mdm_background}
We adopt masked diffusion~\cite{D3PM,LLaDa,ye2025dream} to generate the $n$-digit SID. The forward process applies an absorbing-state mask corruption that replaces a subset of digits with \texttt{[MASK]} under a time-dependent schedule~\cite{DBLP:conf/nips/CampbellBBRDD22,DBLP:conf/nips/ShiHWDT24,DBLP:conf/nips/SahooASGMCRK24}. The reverse process predicts all masked digits in parallel from the corrupted sequence.

We train the decoder with a masked-digit cross-entropy as:
\begin{equation}
\mathcal{L}(\theta)
=
-\,\mathbb{E}_{\mathbf{x}_0,\,\tau,\,\mathbf{x}_\tau \sim q_{\tau\mid 0}(\cdot\mid \mathbf{x}_0)}\!\left[
\frac{1}{|\mathcal{M}_\tau|}\!
\sum_{k\in \mathcal{M}_\tau}
\log p_{\theta}\!\left(x_0^{\,k} \mid \mathbf{x}_\tau, \tau\right)
\right],
\end{equation}
where $\mathbf{x}_0$ is the clean SID sequence, $\mathbf{x}_\tau$ is its corrupted version with mask ratio $\tau\in[0,1)$, and $\mathcal{M}_\tau$ is the masked index set. This objective upper-bounds the negative log-likelihood~\cite{DBLP:conf/nips/ShiHWDT24,DBLP:conf/iclr/OuNXZSLL25}, removes causal constraints, and yields richer supervision signals via discrete masking during training, enabling parallel generation.

\section{Method}\label{sec:method}

\subsection{Overview}\label{subsec:overview}
Figure~\ref{fig:framework} illustrates DiffGRM, which predicts the next item’s SID in parallel via masked diffusion. We describe parallel semantic encoding (PSE) for SID tokenization (Section~\ref{subsec:pse}), introduce on-policy coherent noising (OCN) for difficulty-aware masking (Section~\ref{subsec:ocn}), and present confidence-guided parallel denoising (CPD) for inference (Section~\ref{subsec:cpd}). Finally, we discuss design choices and analyze training and inference complexity (Section~\ref{subsec:discussion}).

We adopt an encoder–decoder architecture, where the encoder summarizes the user history and the masked-diffusion decoder (MD-Decoder) predicts SID digits in parallel under non-causal attention, and the workflow that connects these components is as follows.
Items are first tokenized into SIDs via PSE (Section~\ref{subsec:pse}), which converts each user's interaction history into a sequence of SIDs.
We embed the $n$ digits of each item, concatenate and project them into an item vector, add positional embeddings, and feed the sequence to a Transformer encoder to obtain $\mathbf{H}_u\!\in\!\mathbb{R}^{L_{\text{input}}\times d_m}$, where $L_{\text{input}}$ is the encoder input length and $d_m$ the model hidden size.
The MD-Decoder then takes a partially masked $n$-digit input $\mathbf{y}$ and applies non-causal (bidirectional) self-attention across digits.
Unlike autoregressive causal attention, each digit attends to all others, enabling bidirectional intra-item semantics and cross-digit mutual verification. 
The MD-Decoder then cross-attends to $\mathbf{H}_u$ (encoder-side $\boldsymbol{k}/\boldsymbol{v}$ are derived from $\mathbf{H}_u$) and predicts all masked digits in parallel. During multi-view training, we compute $\mathbf{H}_u$ once per sample and cache encoder-side $\boldsymbol{k}/\boldsymbol{v}$ for reuse across views and CPD steps, which amortizes cross-attention cost since only the decoder queries are recomputed.
For decoding, we adopt confidence-guided parallel denoising in Section~\ref{subsec:cpd} which performs global parallel beam search to complete SIDs.

We train the MD-Decoder with masked-digit cross-entropy on each masked input, called a view. Views are constructed by OCN as described in Section~\ref{subsec:ocn}. For view $r$ with masked index set $\mathcal{M}^{(r)}$ and per-digit distribution $p_{\theta}^{(r,k)}(\cdot \mid \mathbf{y}^{(r)}, \mathbf{H}_u)$ at digit $k$, the per-view objective is given as:
\begin{equation}
\mathcal{L}^{(r)}
=\frac{1}{|\mathcal{M}^{(r)}|}\sum_{k\in\mathcal{M}^{(r)}}
\left(-\sum_{v=0}^{M-1}\tilde{q}^{(k)}_v\,\log p^{(r,k)}_{\theta}\!\left(v \mid \mathbf{y}^{(r)}, \mathbf{H}_u\right)\right),
\end{equation}
where $\tilde{\mathbf{q}}^{(k)}$ is the smoothed one-hot target for digit $k$, $\mathbf{y}^{(r)}$ is the partially masked input of view $r$, and $\mathcal{M}^{(r)}$ specifies which digits are masked. The total loss averages the per-view objectives across a small set of views constructed by OCN.

\subsection{Parallel Semantic Encoding}\label{subsec:pse}
In recommender systems, each item $i \in \mathcal{I}$ typically contains multiple content features $\mathbf{f}_i$, such as textual descriptions, videos, and images. A common practice is to employ a semantic encoder $E(\cdot)$ (e.g., a pretrained language model such as Sentence-T5~\cite{t5} or BERT~\cite{BERT}) to map the raw features into a $d$-dimensional continuous representation $\mathbf{h}_i \in \mathbb{R}^d$. To obtain higher-quality representations, some studies~\cite{LLM2Rec,LLMEmb,EasyRec,BLAIR} further incorporate collaborative signals at this stage to fuse user--item interaction information.  

Once the continuous representation $\mathbf{h}_i$ is obtained, we discretize it into a SID via \emph{Parallel Semantic Encoding} (PSE). We adopt an OPQ-based partition-and-quantize scheme: an orthogonal rotation $\mathbf{R}_o$ is learned to reduce downstream quantization distortion, the rotated vector $\tilde{\mathbf{h}}_i=\mathbf{R}_o\mathbf{h}_i$ is evenly partitioned into $n$ subvectors $\tilde{\mathbf{h}}_i=\mathbf{v}_i^{1}\oplus\cdots\oplus\mathbf{v}_i^{n}$, and each subvector is assigned independently to a per-digit codebook $\mathbf{C}^{(k)}=\{\mathbf{c}^{(k)}_{0},\dots,\mathbf{c}^{(k)}_{M-1}\}$ by nearest-centroid indexing $s_i^{k}=\arg\min_j\|\mathbf{v}_i^{k}-\mathbf{c}^{(k)}_{j}\|_2^2$. This yields an $n$-digit $\mathrm{SID}_i=[s_i^{0},\dots,s_i^{n-1}]$ with decoupled per-digit assignments, which removes residual sequential dependence and enables fully parallel prediction across digits. A rationale for choosing PSE over residual tokenizers is discussed in Section~\ref{subsec:discussion-rq}.

\subsection{On-policy Coherent Noising}\label{subsec:ocn}
Random masking yields sparse and imbalanced supervision per sample. In masked diffusion, covering diverse target–context signals demands many masking patterns, which inflates sample requirements and increases the training burden. See Appendix~\ref{app:sup_details} for a count of supervision signals and sample requirements, where with an $n$-layer codebook the autoregressive paradigm provides $n$ signals that a single teacher-forced sample already covers, whereas masked diffusion admits $n\,2^{\,n-1}$ distinct target--context signals and full coverage would require at least $2^{\,n}-1$ masking configurations.
On-policy coherent noising (OCN) selects mask sets with the current model as a policy $\pi_{\theta}$ over digits. We call each masked input a view, and OCN constructs a small nested set of views per sample ordered from light to heavy corruption. By reallocating the masking budget to hard digits and increasing corruption along the uncertainty order, OCN focuses supervision where it matters most and improves training efficiency.

As shown in Figure~\ref{fig:framework}(a) right, given cached encoder-side $\boldsymbol{k}/\boldsymbol{v}$, we run the MD-Decoder once on a fully masked $n$-digit input, i.e., the last view $R$ with $m_R=n$ and $t_R=1$. This probe yields, for each digit $k$, a predictive distribution $p^{(R,k)}_{\theta}(\cdot)$ over $\{0,\ldots,M-1\}$, and we quantify confidence and difficulty as:
\[
p_{\max}^{(k)}=\max_{v\in\{0,\ldots,M-1\}} p^{(R,k)}_{\theta}(v), \qquad
\delta^{(k)}=1-p_{\max}^{(k)},
\]
where $p_{\max}^{(k)}$ is the model confidence for digit $k$ and $\delta^{(k)}$ measures its difficulty (see the per-digit bars and confidence curve in Figure~\ref{fig:framework}(a)). Larger $\delta^{(k)}$ indicates lower confidence or higher perplexity. These scores induce policy $\pi_{\theta}(k)\propto \delta^{(k)}$. In practice, for view $r$ we deterministically mask the top $m_r$ digits ranked by $\delta^{(k)}$, and a stochastic variant samples $m_r$ digits without replacement under $\pi_{\theta}$.

Building on the difficulty scores, we sort digits by $\delta^{(k)}$ in descending order to obtain a permutation $\sigma$ from hardest to easiest. With a nondecreasing schedule $1\le m_1<\cdots<m_R\le n$, view $r$ masks the $m_r$ hardest digits and keeps the others visible with clean embeddings. The layer-0 input for view $r$ is constructed as:
\begin{equation}
\mathbf{y}^{(r,k)}=
\begin{cases}
\mathbf{E}_{\text{mask}}[k], & \text{if } k\in\mathcal{M}^{(r)},\\[3pt]
\mathbf{E}^{(k)}_{\mathrm{sid}}[s^k], & \text{if } k\notin\mathcal{M}^{(r)},
\end{cases}
\end{equation}
and the masked index set is defined as:
\begin{equation}
\mathcal{M}^{(r)}=\{\sigma(1),\ldots,\sigma(m_r)\},
\end{equation}
while collecting all views row-wise yields the binary view matrix:
\begin{equation}
\mathbf{M}_{\mathrm{ocn}}=
\begin{bmatrix}
(\mathbf{m}^{(1)})^{\top}\\[-2pt]
\vdots\\[-2pt]
(\mathbf{m}^{(R)})^{\top}
\end{bmatrix}\in\{0,1\}^{R\times n},
\end{equation}
where $m_k^{(r)}=\mathbf{1}[k\in\mathcal{M}^{(r)}]$ and $\mathbf{m}^{(r)}=(m_0^{(r)},\ldots,m_{n-1}^{(r)})^{\top}$ encode which digits are masked in view $r$, $\sigma$ orders digits by $\delta^{(k)}$ from hardest to easiest, and $m_r$ is the number of masked digits in view $r$. The masked sets are nested, so the visible context grows across views and the corruption ratio $t_r=m_r/n$ increases from light to heavy masking, which avoids combinatorial masking patterns, stabilizes optimization with progressively richer evidence, and concentrates gradients on the same hard digits under increasing context.

For each view we apply masked-digit cross-entropy on its masked indices and aggregate across views as:
\begin{equation}
\mathcal{L}=\frac{1}{R}\sum_{r=1}^{R}\mathcal{L}^{(r)},
\end{equation}
where $\mathcal{L}^{(r)}$ is computed on $\mathcal{M}^{(r)}$ for view $r$, $R$ is the number of views, and $1\le R\le n$. In practice we compute the difficulty order once per example with a fully masked pass, reuse the encoder output across $R$ views, and break ties in $\delta^{(k)}$ with a fixed digit order.

\subsection{Confidence-guided Parallel Denoising}\label{subsec:cpd}
As shown in Figure~\ref{fig:framework}(b), when $B_{\text{act}}=1$, CPD runs a single branch. Starting at $t_R=1$ from a fully masked input with encoder-side $\boldsymbol{k}/\boldsymbol{v}$ cached, each reverse step selects the most confident digit–codeword pair and fills that digit while remasking the others.

Unlike LLM-oriented discrete MDMs that use greedy fill-in and yield only a single Top-1 output~\cite{DBLP:conf/nips/AustinJHTB21,LLaDa}, recommendation systems must return a diverse Top-$K$ candidate set to support downstream ranking and serving. In the GR framework this need is typically met by autoregressive decoders with left-to-right beam search that expand one digit at a time. To fully exploit the MDM’s parallel generation capacity, we depart from left-to-right expansion and propose Confidence-guided Parallel Denoising (CPD), a global beam-search denoiser on the MD-Decoder that jointly scores partial SIDs, fills digits in descending model confidence, and yields the Top-$K$ candidates. We next generalize CPD to beam width $B_{\text{act}}>1$ as follows.

Formally, for beam width $B_{\text{act}}>1$, CPD proceeds over reverse steps $\{t_r\}_{r=1}^{R}$ with $t_R=1$ and $t_1=0$, maintaining an active set of partial SIDs; the encoder-side $\boldsymbol{k}/\boldsymbol{v}$ are computed once and cached for all steps. At $t_R=1$ with the fully masked input $\mathbf{y}^{(R)}$, we initialize the active set as:
\begin{equation}
\mathcal{B}_{R}
= \operatorname*{Top}_{B_{\text{act}}}\!\Big\{\log p_{\theta}\!\left(y_k{=}c \,\middle|\, \mathbf{y}^{(R)}, \mathbf{H}_u\right)\Big\},
\end{equation}
where $y_k$ is the $k$-th SID digit, $k\in\{0,\ldots,n-1\}$, $c\in\{0,\ldots,M-1\}$, $M$ is the per-digit codebook size, $B_{\text{act}}$ is the per-step beam width, $p_{\theta}$ is the MD-Decoder distribution, and $\operatorname*{Top}_{B_{\text{act}}}\{\cdot\}$ selects the top $B_{\text{act}}$ scores across all $(k,c)$ pairs.

At denoising step $t_r$ we score filling one still-masked digit of each active branch as:
\begin{equation}
s_{r-1}(b,k,c)=\mathrm{score}_{r}(b)+\log p_{\theta}\!\left(y_k{=}c \mid \mathbf{y}^{(r)}_{b}, \mathbf{H}_u\right),
\end{equation}
where $b\!\in\!\mathcal{B}_{r}$ indexes a branch, $\mathrm{score}_{r}(b)$ is its accumulated log-probability at $t_r$, $\mathbf{y}^{(r)}_{b}$ is its partially denoised SID at $t_r$, and $k\!\in\!\mathcal{M}^{(b)}_{r}$ is a still-masked index. Per-step truncation keeps the strongest children and advances the reverse schedule as:
\begin{equation}
\mathcal{B}_{r-1}
= \operatorname*{Top}_{B_{\text{act}}}\!\Big\{\, s_{r-1}(b,k,c)\,\Big\},
\end{equation}
where $b\in\mathcal{B}_{r}$, $k\in\mathcal{M}^{(b)}_{r}$, $c\in\{0,\ldots,M-1\}$, and $\operatorname*{Top}_{B_{\text{act}}}\{\cdot\}$ returns the $B_{\text{act}}$ highest-scoring elements. After selecting $\mathcal{B}_{r-1}$ we fill the chosen digits for each tuple $(b,k,c)$ and remove $k$ from the masked-index set of branch $b$ while all other digits remain masked, so the mask ratio decreases from $t_r$ to $t_{r-1}$. The iteration ends at $t_1=0$ yielding $\mathcal{B}_{0}$, after which we deduplicate the generated sequences and keep the Top-$K$ by final scores.

\begin{table}[t]
\centering
\caption{Asymptotic complexity per training sample and per inference request, showing leading terms only. The encoder is computed once and shared. At inference both paradigms include the common encoder term \(O(N^{2}d_m)\). The balance between encoder and decoder depends on \(N\) relative to \(n B_{\text{act}}\). DiffGRM adds one extra factor \(n\) in the decoder term.}
\label{tab:complexity}
\setlength{\tabcolsep}{4pt}
\begin{tabular}{lcc}
\toprule
\textbf{Module} & \textbf{ARM} & \textbf{DiffGRM} \\
\midrule
Encoder & \(O(N^{2}d_m)\) & \(O(N^{2}d_m)\) \\
Decoder—training & \(O(n^{2}d_m + n N d_m)\) & \(R \cdot O(n^{2}d_m + n N d_m)\) \\
Decoder—inference & \(O(B_{\text{act}}\, n\, N d_m)\) & \(O(B_{\text{act}}\, n^{2}\, N d_m)\) \\
\bottomrule
\end{tabular}
\end{table}

\subsection{Discussion}\label{subsec:discussion}

\subsubsection{Why not RQ}\label{subsec:discussion-rq}
RQ is widely used in generative recommendation tokenizers (e.g., RQ-VAE, RQ-Kmeans), with 3--4 codebook layers~\cite{TIGER,GRID,OneRec,OneRec_v1}.
In our setting, items are $n$-digit SIDs and the generator predicts digits in parallel with diffusion, but RQ has two drawbacks.
First, residual levels often have unbalanced information, exacerbating inter-digit heterogeneity and uneven predictability~\cite{DBLP:journals/corr/abs-2410-09560}.
Second, hierarchical residual dependence couples later digits to earlier ones, introducing left-to-right bias that conflicts with parallel, any-order prediction~\cite{SETRec,RPG}.
In contrast, PSE factorizes representations into independent subspaces (e.g., OPQ~\cite{OPQ}, FSQ~\cite{FSQ}), balancing per-digit information and removing sequential coupling, which better matches diffusion-based generation.

\subsubsection{Complexity of Training}
We analyze an encoder--decoder setup where the encoder processes a history of length \(N\) and the decoder predicts an \(n\)-digit SID. Let the model width be \(d_m\), codebook size \(M\), and \(R\) the number of coherent views. For \textbf{ARM} with teacher forcing, the encoder costs \(O(N^{2}d_m + N d_m^{2})\) and one decoder pass contributes \(O(n^{2}d_m + n N d_m + n d_m^{2} + n M d_m)\). The dominant per-sample step is \(O(N^{2}d_m) + O(n^{2}d_m + n N d_m)\). For \textbf{DiffGRM}, one encoder run is reused across \(R\) views, giving \(O(N^{2}d_m) + R \cdot O(n^{2}d_m + n N d_m)\). See Table~\ref{tab:complexity} for leading orders. In industrial settings, histories are long and $R$ is small; training is encoder-dominated, so ARM and DiffGRM have essentially the same overall training complexity.

\begin{table}[b]
\centering
\caption{Statistics of the processed datasets. ``Avg.\ $t$'' denotes the average number of interactions per input sequence.}
\label{tab:data_stats}
\begin{tabular}{lcccc}
\toprule
\textbf{Dataset} & \textbf{\#Users} & \textbf{\#Items} & \textbf{\#Interactions} & \textbf{Avg.\ $t$} \\
\midrule
\textbf{Sports} & 35{,}598 & 18{,}357 & 260{,}739 & 8.32 \\
\textbf{Beauty} & 22{,}363 & 12{,}101 & 176{,}139 & 8.87 \\
\textbf{Toys}   & 19{,}412 & 11{,}924 & 148{,}185 & 8.63 \\
\bottomrule
\end{tabular}
\end{table}

\begingroup
\renewcommand{\arraystretch}{0.95} % 默认 1；>1 行更高，<1 行更紧
\begin{table*}[t]
\centering
\caption{Overall performance on Amazon Reviews (Sports, Beauty, Toys) by Recall@K and NDCG@K ($K\!\in\!\{5,10\}$). Best results are in \textbf{bold}; second-best are \underline{underlined}. \textbf{DiffGRM} significantly outperforms the strongest baseline (paired t-test, $p<0.05$). “—” denotes results not reported by the original paper, and other baseline numbers are taken from prior publications~\cite{RPG,ActionPiece,TIGER}.
}
\label{tab:overall-perf}
\begin{tabularx}{\textwidth}{@{}l*{12}{Y}@{}}
\toprule
\multirow{3}{*}{\textbf{Methods}} &
\multicolumn{4}{c}{\textbf{Sports and Outdoors}} &
\multicolumn{4}{c}{\textbf{Beauty}} &
\multicolumn{4}{c}{\textbf{Toys and Games}} \\
\cmidrule(lr){2-5}\cmidrule(lr){6-9}\cmidrule(lr){10-13}
& \shortstack{Recall\\@5} & \shortstack{NDCG\\@5} & \shortstack{Recall\\@10} & \shortstack{NDCG\\@10}
& \shortstack{Recall\\@5} & \shortstack{NDCG\\@5} & \shortstack{Recall\\@10} & \shortstack{NDCG\\@10}
& \shortstack{Recall\\@5} & \shortstack{NDCG\\@5} & \shortstack{Recall\\@10} & \shortstack{NDCG\\@10} \\

\midrule\midrule
\multicolumn{13}{c}{\emph{Item ID-based (Discriminative)}} \\
\midrule

GRU4Rec     & 0.0129 & 0.0086 & 0.0204 & 0.0110 & 0.0164 & 0.0099 & 0.0283 & 0.0137 & 0.0097 & 0.0059 & 0.0176 & 0.0084 \\
HGN         & 0.0189 & 0.0120 & 0.0313 & 0.0159 & 0.0325 & 0.0206 & 0.0512 & 0.0266 & 0.0321 & 0.0221 & 0.0497 & 0.0277 \\
SASRec      & 0.0233 & 0.0154 & 0.0350 & 0.0192 & 0.0387 & 0.0249 & 0.0605 & 0.0318 & 0.0463 & 0.0306 & 0.0675 & 0.0374 \\
BERT4Rec    & 0.0115 & 0.0075 & 0.0191 & 0.0099 & 0.0203 & 0.0124 & 0.0347 & 0.0170 & 0.0116 & 0.0071 & 0.0203 & 0.0099 \\
\midrule
\multicolumn{13}{c}{\emph{Semantic-enhanced (Discriminative)}} \\
\midrule
FDSA        & 0.0182 & 0.0122 & 0.0288 & 0.0156 & 0.0267 & 0.0163 & 0.0407 & 0.0208 & 0.0228 & 0.0140 & 0.0381 & 0.0189 \\
S$^3$-Rec   & 0.0251 & 0.0161 & 0.0385 & 0.0204 & 0.0387 & 0.0244 & 0.0647 & 0.0327 & 0.0443 & 0.0294 & 0.0700 & 0.0376 \\
VQ-Rec      & 0.0208 & 0.0144 & 0.0300 & 0.0173 & 0.0457 & 0.0317 & 0.0664 & 0.0383 & 0.0497 & 0.0346 & 0.0737 & 0.0423 \\
RecJPQ      & 0.0141 & 0.0076 & 0.0220 & 0.0102 & 0.0311 & 0.0167 & 0.0482 & 0.0222 & 0.0331 & 0.0182 & 0.0484 & 0.0231 \\
\midrule
\multicolumn{13}{c}{\emph{Semantic ID-based (Generative)}} \\
\midrule
TIGER       & 0.0264 & 0.0181 & 0.0400 & 0.0225 & 0.0454 & 0.0321 & 0.0648 & 0.0384 & 0.0521 & 0.0371 & 0.0712 & 0.0432 \\
HSTU        & 0.0258 & 0.0165 & 0.0414 & 0.0215 & 0.0469 & 0.0314 & 0.0704 & 0.0389 & 0.0433 & 0.0281 & 0.0669 & 0.0357 \\
ActionPiece & \underline{0.0316} & 0.0205 & \underline{0.0500} & \underline{0.0264} & 0.0511 & 0.0340 & 0.0775 & 0.0424 & — & — & — & — \\
RPG         & 0.0314 & \underline{0.0216} & 0.0463 & 0.0263 & \underline{0.0550} & \underline{0.0381} & \underline{0.0809} & \underline{0.0464} & \underline{0.0592} & \underline{0.0401} & \textbf{0.0869} & \underline{0.0490} \\

\midrule

\textbf{DiffGRM} & \textbf{0.0363} & \textbf{0.0245} & \textbf{0.0550} & \textbf{0.0305} & \textbf{0.0603} & \textbf{0.0414} & \textbf{0.0876} & \textbf{0.0502} & \textbf{0.0618} & \textbf{0.0455} & \underline{0.0834} & \textbf{0.0524} \\

\rowcolor{RowBlue}
\textbf{Improv.} & +14.87\% & +13.43\% & +10.00\% & +15.53\% & +9.64\% & +8.66\% & +8.28\% & +8.19\% & +4.39\% & +13.47\% & -4.03\% & 6.94\% \\

\bottomrule
\end{tabularx}

\end{table*}
\endgroup

\subsubsection{Complexity of Inference}
Let \(B_{\text{act}}\) be the active beam width and \(n_r\) the unresolved digits at reverse step \(r\). Both \textbf{ARM} and \textbf{DiffGRM} run the encoder once at \(O(N^{2}d_m + N d_m^{2})\). The ARM decoder performs \(n\) incremental steps, with a leading \(O(B_{\text{act}}\, n\, N d_m)\) per request. DiffGRM with CPD adds a fully masked pass and reverse steps with \(\sum_r n_r=\Theta(n^{2})\), yielding \(O(B_{\text{act}}\, n^{2}\, N d_m)\). When \(B_{\text{act}}\) is large the decoder side can take a noticeable share, yet \(n\) is small and \(N\) is typically the largest quantity, so the common encoder term remains substantial. The relative weight of encoder versus decoder depends on \(N\) compared to \(n B_{\text{act}}\). Overall, DiffGRM and ARM have similar end-to-end inference cost, and the additional factor in the DiffGRM decoder introduces only a modest overhead. Because the extra scoring runs in parallel across digits and beams using the cached encoder keys and values, it increases compute but not the critical path, so the wall-clock latency is typically close to ARM.

\section{Experiments}
We evaluate DiffGRM and focus on these research questions:

\noindent\textbf{RQ1:} How does DiffGRM perform compared with strong discriminative and generative recommendation baselines? 

\noindent\textbf{RQ2:} Under a limited training budget, does OCN allocate supervision more effectively and improve sample efficiency than random masking or fixed coherent-path masking?

\noindent\textbf{RQ3:} How do the key components contribute individually and jointly to overall performance?

\subsection{Experimental Setup}

\noindent\textbf{Datasets.}
We evaluate on three Amazon Reviews categories~\cite{DBLP:conf/sigir/McAuleyTSH15}:
``Sports and Outdoors'' (\textbf{Sports}), ``Beauty'' (\textbf{Beauty}), and ``Toys and Games'' (\textbf{Toys}),
which are standard benchmarks for semantic ID--based generative recommendation~\cite{TIGER,LMIndexer,P5-CID}.
Following prior work~\cite{TIGER,VQ-Rec,S3-Rec}, we treat each user's historical reviews as interactions and sort them chronologically to form sequences. We adopt the standard leave-last-out evaluation~\cite{SASRec,TIGER,DBLP:journals/tois/ZhaoLFWW23}: the last item in each sequence is used for testing, the second-to-last for validation, and the remaining interactions for training. Table~\ref{tab:data_stats} summarizes dataset statistics.

\begin{figure*}[t]
  \centering
  \includegraphics[width=\textwidth]{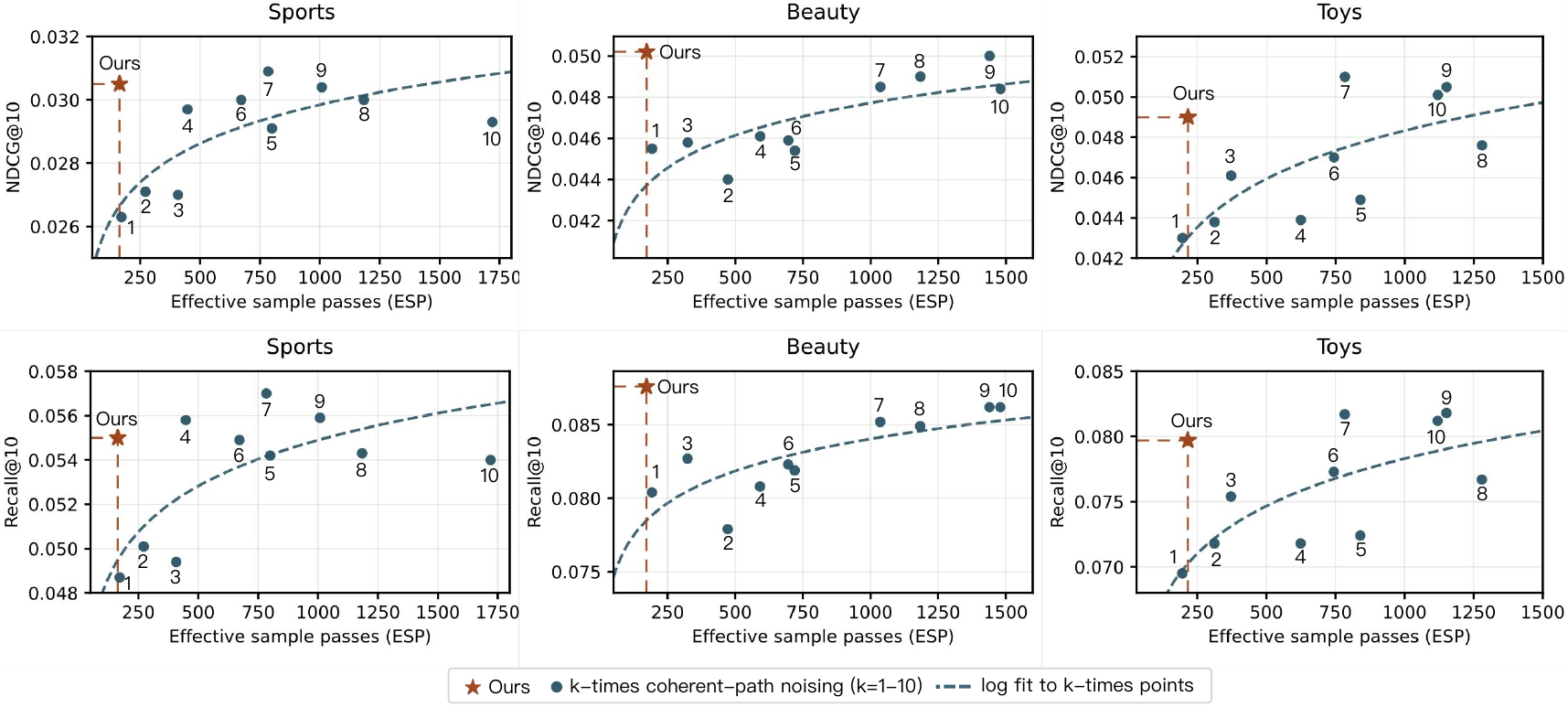}
  \caption{Analysis of performance (NDCG@10/Recall@10) w.r.t.\ effective sample passes (ESP). DiffGRM matches or surpasses the $k$-times coherent-path noising settings under lower ESP, indicating higher sample efficiency. For fairness, on Toys we fix $d_{\text{model}}{=}256$, while the optimal $d_{\text{model}}{=}1024$ runs out of memory for $k{>}3$.}
  \label{fig:exp_ocn}
\end{figure*}

\noindent\textbf{Baselines.}
We compare DiffGRM with the following baselines, grouped into three families: (i) Item ID-based discriminative models, (ii) Semantic-enhanced discriminative models, and (iii) Semantic ID-based generative models.

\begin{itemize}[leftmargin=*,nosep]
  \item \textbf{GRU4Rec}~\cite{GRU4Rec}: an RNN-based session recommender that models sequential dynamics.
  \item \textbf{HGN}~\cite{HGN}: gating improves RNN-based sequence modeling.
  \item \textbf{SASRec}~\cite{SASRec}: self-attentive Transformer decoder with binary cross-entropy for next-item prediction.
  \item \textbf{BERT4Rec}~\cite{BERT4Rec}: bidirectional Transformer encoder trained with a Cloze-style objective on item IDs.
  \item \textbf{FDSA}~\cite{FDSA}: models and fuses item-ID and feature sequences with self-attention.
  \item \textbf{S$^3$-Rec}~\cite{S3-Rec}: self-supervised pretraining on features and IDs, then fine-tuning for next-item prediction.
  \item \textbf{VQ-Rec}~\cite{VQ-Rec}: product-quantizes text features into semantic IDs and pools them as item representations.
  \item \textbf{RecJPQ}~\cite{RecJPQ}: replaces item embeddings with concatenated jointly product-quantized sub-embeddings.
  \item \textbf{TIGER}~\cite{TIGER}: RQ-VAE tokenization and autoregressive generation of the next SID token.
  \item \textbf{HSTU}~\cite{HSTU}: discretizes raw item features as tokens for generative recommendation; for consistency with prior setups~\cite{RPG}, we adopt 4-digit OPQ-tokenized SIDs as item tokens.
  \item \textbf{ActionPiece}~\cite{ActionPiece}: context-aware tokenization that represents each action as an unordered set of item features.
  \item \textbf{RPG}~\cite{RPG}: predicts unordered SID tokens in parallel via a multi-token objective with graph-guided decoding.
\end{itemize}

\noindent\textbf{Evaluation settings.}
We use Recall@$K$ and NDCG@$K$ as metrics to evaluate the methods, where $K\in\{5,10\}$, following Rajput et al.~\cite{TIGER}.
Model checkpoints with the best performance on the validation set are used for evaluation on the test set.

\noindent\textbf{Implementation details.}
Please refer to Appendix~\ref{app:impl} for detailed implementation and hyperparameter settings.

\subsection{Overall Performance (RQ1)}
We compare DiffGRM with both discriminative and generative baselines on three benchmarks using SID-based evaluation, as summarized in Table~\ref{tab:overall-perf}. Methods that leverage semantic information consistently outperform ID-only discriminative models, and within the semantic family, semantic ID-based generative models generally surpass semantic-enhanced discriminative ones.

DiffGRM achieves the best overall results, ranking first on 11 of the 12 metrics. Relative to the strongest baseline, NDCG at K=10 improves by 15.53\% on Sports, 8.19\% on Beauty, and 6.94\% on Toys. Recall at K=10 improves by 10.00\% on Sports and 8.28\% on Beauty, and is slightly lower on Toys by 4.03\%, while DiffGRM still attains higher NDCG at both K=5 and K=10. These gains stem from masked-diffusion training over SIDs that supplies dense per-digit supervision and bidirectional context among SID digits, together with OCN that allocates more supervision to difficult digits and CPD that enables parallel Top-$K$ SID generation. We report results with alternative semantic encoders in Appendix~\ref{app:expressive}.

\subsection{Effectiveness of OCN (RQ2)}
This section evaluates whether On-policy Coherent Noising (OCN) achieves a more balanced supervision allocation under a smaller training budget. Analysis of supervision-signal differences between the ARM and the MDM is provided in Appendix~\ref{app:sup_details}.

We introduce a unified measure, effective sample passes (ESP): \(\mathrm{ESP}=\texttt{best\_epoch}\times\) (training views per sample per epoch). Training views come from coherent-path noising: with one path, each sample yields \(n\) views per epoch, equal to the number of codebook layers (\(n=4\)). To test whether more supervision helps, we expand the coherent paths from 1 to \(k\), so views become \(n\times k\) and \(\mathrm{ESP}=\texttt{best\_epoch}\times n\times k\) (for fairness, we apply the same view expansion to ARM; see Appendix~\ref{app:arm_data_augmentation}). ESP is independent of the number of training samples (per-dataset numbers of training samples and sliding-window augmentation details are in Appendix~\ref{app:data_augmentation}).

As shown in Figure~\ref{fig:exp_ocn}, increasing \(k\) raises performance but also raises ESP. The dashed line is a logarithmic least-squares fit over the \(k\)-times coherent-path points, summarizing how performance scales with ESP. In contrast, On-policy Coherent Noising (OCN) achieves better results at the same or lower ESP by using the current model to select the most uncertain positions along coherent paths, focusing training on high-value signals and improving sample efficiency.

\begin{table}[t]
\centering
\caption{Ablation analysis of DiffGRM. The recommendation
performance is measured using NDCG@10. The best per-
formance is denoted in bold fonts.}
\label{tab:ablation}
\begin{tabular}{@{}lccc@{}}
\toprule
\textbf{Variants} & \textbf{Sports} & \textbf{Beauty} & \textbf{Toys} \\
\midrule
\multicolumn{4}{c}{\emph{Semantic ID Setting}} \\
\midrule
(1.1) PSE $\rightarrow$ RQ-Kmeans & 0.0200 & 0.0343 & 0.0305 \\
(1.2) PSE $\rightarrow$ Random    & 0.0138 & 0.0300 & 0.0206 \\
\midrule
\multicolumn{4}{c}{\emph{Training strategy}} \\
\midrule
(2.1) \emph{w/o} OCN & 0.0250 & 0.0368 & 0.0385 \\
(2.2) \emph{w/o} On-policy & 0.0263 & 0.0455 & 0.0430 \\
\midrule
\multicolumn{4}{c}{\emph{Inference strategy}} \\
\midrule
(3.1) \emph{w/o} CPD & 0.0273 & 0.0496 & 0.0499 \\
\midrule
DiffGRM (ours) & \textbf{0.0305} & \textbf{0.0502} & \textbf{0.0524} \\
\bottomrule
\end{tabular}
\end{table}

\subsection{Ablation Study (RQ3)}
We conduct ablation analyses in Table~\ref{tab:ablation} to quantify each module's contribution to DiffGRM's overall performance.

(1) To assess Parallel Semantic Encoding, we compare two variants. (1.1) Replacing PSE with RQ-KMeans degrades performance because the residual hierarchy in RQ introduces dependencies across digits that conflict with masked diffusion's bidirectional and parallel denoising. (1.2) Replacing PSE with random tokens further hurts by discarding semantic structure. This confirms that PSE is necessary for MDM.

(2) Then, to assess On-policy Coherent Noising, we compare two variants. (2.1) \emph{w/o} OCN adopts DDMs-style random masking~\cite{LLaDa}, removing coherent noising and on-policy selection, which disperses supervision and leaves infrequent long-tail samples insufficiently trained. (2.2) \emph{w/o} on-policy keeps coherent noising but removes on-policy selection, performing better than (2.1) yet still below DiffGRM. This shows that on-policy further focuses supervision on weak positions.

(3) Finally, to assess Confidence-guided Parallel Denoising, \emph{w/o} CPD replaces confidence-guided parallel denoising with random fixed-order beam search, which decodes digits in a fixed random permutation without confidence feedback and reduces performance across all three datasets.

\subsection{Further Analysis}

\subsubsection{OCN Strategy Analysis}

\begin{table}[t]
\centering
\caption{OCN variants on \textbf{Beauty} and \textbf{Toys} (NDCG@10). \textbf{Improv.}: \((\emph{w/o}\,\text{CPD}-\text{CPD})/\text{CPD}\). Abbreviations: \textbf{L-S} (least, static), \textbf{L-R} (least, refresh), \textbf{M-S} (most, static), \textbf{M-R} (most, refresh).}
\label{tab:ocn_variants}
\setlength{\tabcolsep}{4pt}
\begin{tabular}{llcccc}
\toprule
\textbf{Dataset} & \textbf{Metric} & \textbf{L-S (Ours)} & \textbf{L-R} & \textbf{M-S} & \textbf{M-R} \\
\midrule
\multirow{3}{*}{Beauty}
& CPD            & 0.0502 & 0.0484 & 0.0476 & 0.0382 \\
& \emph{w/o} CPD & 0.0496 & 0.0470 & 0.0444 & 0.0309 \\
& \cellcolor{RowBlue}Improv.
                  & \cellcolor{RowBlue}\(-1.20\%\) & \cellcolor{RowBlue}\(-2.89\%\) & \cellcolor{RowBlue}\(-6.72\%\) & \cellcolor{RowBlue}\(-19.11\%\) \\
\midrule
\multirow{3}{*}{Toys}
& CPD            & 0.0524 & 0.0481 & 0.0516 & 0.0421 \\
& \emph{w/o} CPD & 0.0499 & 0.0455 & 0.0506 & 0.0318 \\
& \cellcolor{RowBlue}Improv.
                  & \cellcolor{RowBlue}\(-4.71\%\) & \cellcolor{RowBlue}\(-5.41\%\) & \cellcolor{RowBlue}\(-1.94\%\) & \cellcolor{RowBlue}\(-24.47\%\) \\
\bottomrule
\end{tabular}
\end{table}

We compare four OCN variants along two dimensions. \emph{Selection policy} decides which digits are noised earlier: “least” selects the lowest-confidence digits under the current model, while “most” selects the highest-confidence digits. \emph{Refresh frequency} specifies how the order is obtained: “static” runs the MD-Decoder once to estimate uncertainties and then keeps the order fixed for the whole example, while “refresh” re-estimates uncertainties after each denoising step and updates the order at every step, which invokes the decoder \emph{n} times for \emph{n} codebook layers. Results are shown in Table~\ref{tab:ocn_variants}.

\noindent\textbf{Findings.}
(1) \emph{Selection policy}: least-based scheduling outperforms most-based under the same refresh setting.  
(2) \emph{Refresh frequency}: static outperforms refresh; re-estimating the order at each step changes the plan and degrades performance.  
(3) \emph{Order sensitivity}: largest degradation occurs in M-R when replacing CPD with \emph{w/o} CPD. Prioritizing the most confident digits with stepwise refresh makes the model rely on an easy-first order; once the order changes, hard digits are undertrained and performance drops sharply.

\subsubsection{CPD Beam-Size Analysis}

\begin{figure}[t]
  \centering
  \includegraphics[width=\columnwidth]{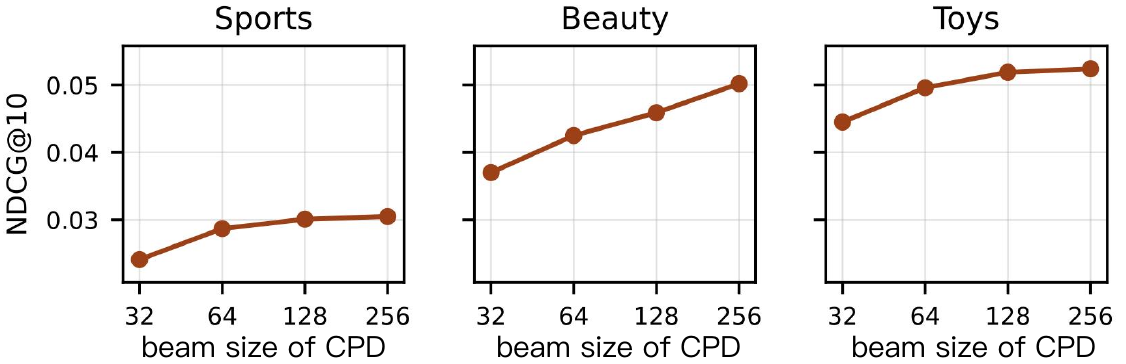}
  \caption{Analysis of DiffGRM performance (NDCG@10) w.r.t.\ beam size in CPD.}
  \label{fig:beam}
\end{figure}

We vary the CPD beam size \{32, 64, 128, 256\} and report results in Figure~\ref{fig:beam}. As in classic beam search, larger beams improve NDCG@10 by mitigating local optima.

\subsubsection{Hidden Dimension Analysis}
Please refer to Appendix~\ref{sec:dim} for analysis of the impact of the hidden dimension $d_m$. In short, $d_m=256$ suits Sports and Beauty; Toys benefits from $d_m=1024$. We use these settings in the main experiments.

\section{Related Work}

\tbf{Generative Recommendation Models.}
Autoregressive (AR) generation casts recommendation as sequence generation~\cite{DBLP:conf/www/RendleFS10,DBLP:conf/www/HouZS0WCM25,SETRec}: items are discretized into semantic IDs and a Transformer predicts the target SID token by token~\cite{SASRec,TIGER,DBLP:conf/icml/JinZ0CW0W0LLW0T24,DBLP:conf/aaai/LiZZYYL25}. 
Representative models include TIGER~\cite{TIGER} mapping items via RQ-VAE and decoding autoregressively; HSTU~\cite{HSTU} framing recommendation as sequence transduction at scale; GenNewsRec~\cite{GenNewsRec} integrating LLM reasoning with item generation; MTGRec~\cite{MTGRec} and ETEGRec~\cite{ETEGRec} enhancing quantization for higher-quality tokens; ActionPiece~\cite{ActionPiece} performing context-aware tokenization; and RPG~\cite{RPG} predicting unordered semantic IDs in parallel.
AR generative recommenders unify representation and prediction, benefit from large-scale language-modeling techniques for stronger sequence modeling~\cite{DBLP:conf/wsdm/TangW18,DBLP:conf/icde/ZhengHLCZCW24}, and support open-vocabulary recommendation~\cite{ActionPiece,DBLP:conf/emnlp/SchmidtRZAUPT24}. 
In summary, left-to-right decoding introduces mismatch and imbalance, motivating diffusion-based alternatives.

\noindent \tbf{Discrete Diffusion Language Models.}
Diffusion models originated for continuous data~\cite{sohl2015deep,song2019generative,ho2020denoising,DBLP:conf/www/Mao0LLLH25}, and were later extended to discrete spaces~\cite{austin2021structured,hoogeboom2021autoregressive}. A continuous-time view models discrete diffusion as a CTMC~\cite{campbell2022continuous}, and concrete-score/score-matching objectives enable effective training~\cite{meng2022concrete,lou2023discrete}. Large-scale discrete diffusion language models (DDMs) now approach autoregressive performance on some tasks~\cite{ye2025dream}, with further gains from improved reverse-sampling strategies~\cite{DBLP:journals/corr/abs-2502-03540,DBLP:conf/iclr/ParkLHTM25,DBLP:journals/corr/abs-2506-19037}. However, these advances target free-form, single-output text, whereas GR requires structured $n$-digit SIDs and a Top-$K$ set. We therefore adapt DDMs to GR with task-specific tokenization, training, and inference.

\section{Conclusion}
In this work, we analyze the structure of semantic ID sequences and reveal the mismatch between autoregressive generation and cross-digit semantics. We present DiffGRM, a diffusion-based generative recommendation framework using discrete masked diffusion. The design combines three components: PSE with OPQ subspace quantization to eliminate cross-digit dependence and balance information; OCN for on-policy coherent noising that allocates supervision by uncertainty and emphasizes difficult digits; and CPD for confidence-guided parallel beam search that produces diverse Top-$K$ candidates. Overall, DiffGRM reconciles cross-digit semantics with parallel generation, yielding state-of-the-art results—improving NDCG@10 by 6.9\%–15.5\% over strong baselines—and producing robust Top-$K$ recommendations. In the future, we will further investigate inference efficiency and scalability.

% \clearpage

\bibliographystyle{ACM-Reference-Format}
\bibliography{references}

% ==== Appendix ====
\appendix

\section{Notations}\label{app:notations}
We summarize the notations used in this paper in Table~\ref{tab:notations_core}.

\begin{table}[b]
\centering
\caption{Notations and explanations.}
\label{tab:notations_core}
\setlength{\tabcolsep}{6pt}
\begin{tabular}{ll}
\toprule
\textbf{Notation} & \textbf{Explanation} \\
\midrule
$\mathrm{SID}_i$ & Item’s semantic ID code. \\
$s_i^k$ & $k$-th SID digit. \\
$n$ & \# codebook layers. \\
$M$ & Per-digit codebook size. \\
$d_m$ & Model hidden dimension. \\
$\alpha$ & Label-smoothing coefficient. \\
$\mathbf{E}_{\mathrm{sid}}^{(k)}$ & SID embedding table (digit--$k$). \\
$\mathbf{E}_{\text{mask}}$ & Mask embedding vector. \\
$\mathbf{H}_u$ & Contextual user-history representation. \\
$L_{\text{input}}$ & Fixed encoder input length. \\
$\mathbf{y},\,\mathbf{y}^{(r)}$ & Partially masked input / view-$r$ sequence. \\
$\tau$ & Mask ratio (time step). \\
$(\mathbf{x}_0,\mathbf{x}_\tau)$ & Clean / corrupted SID sequence. \\
$\mathcal{M}_\tau$ & Masked index set at $\tau$. \\
$R$ & \# coherent views (OCN). \\
$\mathcal{M}^{(r)}$ & Masked indices in view $r$. \\
$(m_r,t_r)$ & Mask count / ratio in view $r$. \\
$\sigma$ & Difficulty order (hard$\rightarrow$easy). \\
$(p_{\max}^{(k)},\delta^{(k)})$ & Confidence / difficulty score. \\
$\mathcal{B}_r$ & Active beam set (CPD). \\
$B_{\text{act}}$ & Per-step beam width (CPD). \\
$K$ & Top-$K$ decoded outputs. \\
\bottomrule
\end{tabular}
\end{table}

\section{Implementation Details}\label{app:impl}
\noindent\textbf{Baselines.}
We use most of the baseline results reported on the same Amazon splits in prior work~\cite{TIGER,ActionPiece,LMIndexer,RPG}.
For the remaining methods, we reproduce them using public implementations—primarily RecBole~\cite{RecBole} or authors’ official code—and tune hyperparameters following the original papers.

\begin{table}[t]
\centering
\caption{Hyperparameters of \textbf{DiffGRM} for each dataset.}
\label{tab:diffgrm_hparams}
\setlength{\tabcolsep}{6pt}
\begin{tabular}{lccc}
\toprule
\textbf{Hyperparameter} & \textbf{Sports} & \textbf{Beauty} & \textbf{Toys} \\
\midrule
learning\_rate        & 0.003 & 0.01  & 0.003 \\
warmup\_steps         & 10{,}000 & 10{,}000 & 10{,}000 \\
dropout\_rate         & 0.1   & 0.1  & 0.1 \\
$d_m$  & 256   & 256  & 1024 \\
d\_ff       & 1024  & 1024 & 1024 \\
num\_heads            & 4     & 4    & 8 \\
$n$                   & 4     & 4    & 4 \\
$M$                   & 256   & 256  & 256 \\
encoder\_layers       & 1     & 1    & 1 \\
md\_decoder\_layers   & 4     & 4    & 4 \\
$\alpha$      & 0.1   & 0.1  & 0.15 \\
$L_{\text{input}}$     & 50   & 50  & 50 \\
$B_{act}$           & 128   & 256  & 128 \\
max\_epochs           & 100   & 100  & 100 \\
early\_stop\_patience & 15    & 15   & 15 \\
\bottomrule
\end{tabular}
\end{table}

\noindent\textbf{DiffGRM.}
We implement our method in PyTorch~\cite{Pytorch} and use FAISS~\cite{FAISS} to OPQ-quantize text-derived item embeddings into SIDs. The backbone is a Transformer encoder~\cite{Transformer} with an MD-Decoder. We adopt dataset-specific hyperparameters summarized in Table~\ref{tab:diffgrm_hparams}. Unless otherwise specified, we use \texttt{sentence-t5-base}~\cite{t5} as the semantic encoder for fairness with prior work, and we report variants with \texttt{bge-large-en-v1.5}~\cite{bge} and \texttt{gte-large-en-v1.5}~\cite{gte} to probe DiffGRM’s ability to capture semantics. We train for at most 100 epochs with AdamW, linear warmup of 10{,}000 steps, dropout 0.1, and label smoothing, and we use early stopping with patience 15 based on the validation score $0.8\,\mathrm{NDCG}@10+0.2\,\mathrm{Recall}@10$. The best checkpoint is used for testing. We decode SID sequences via stepwise masked-diffusion denoising and apply beam search to generate candidates,

\begin{figure*}[t]
  \centering
  \includegraphics[width=\textwidth]{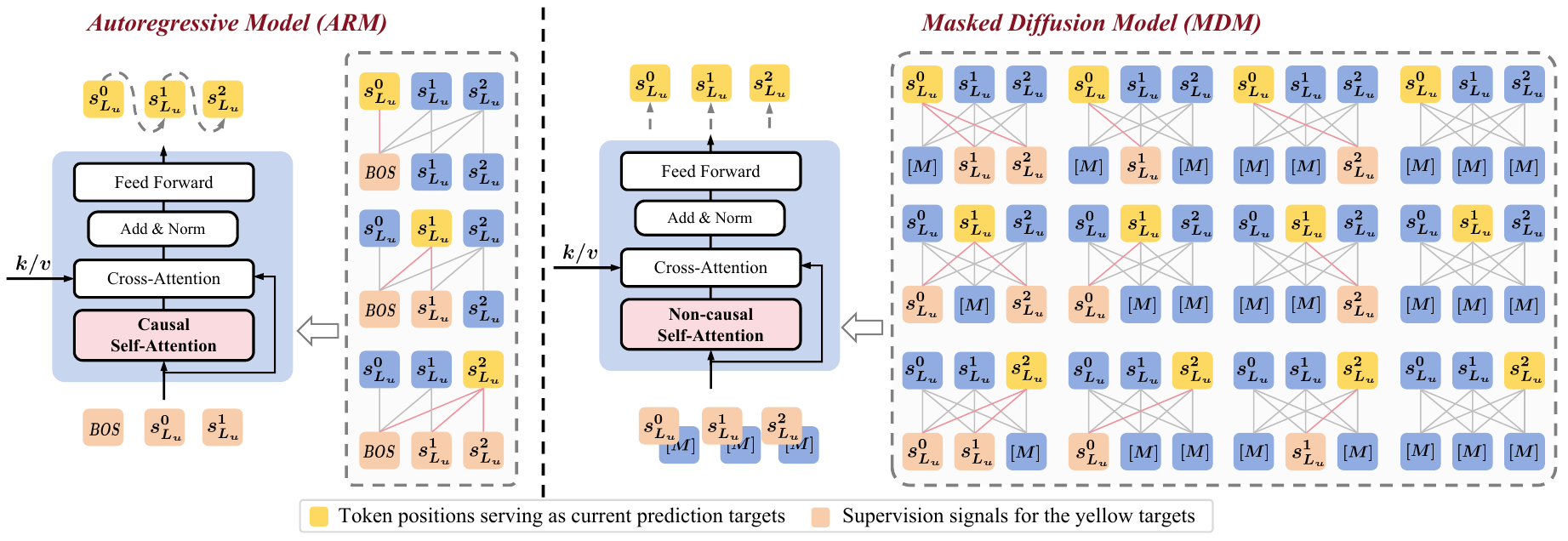}
  \caption{
  Example with a 3-layer codebook (\(n{=}3\)) and semantic ID ($s_{L_u}^{0}, s_{L_u}^{1}, s_{L_u}^{2}$).
  Left: all learnable supervision signals for the autoregressive model (ARM).
  Right: all learnable supervision signals for the masked diffusion model (MDM).}
  \label{fig:sup_signals}
\end{figure*}

\section{Expressive Ability Analysis}\label{app:expressive}
To verify DiffGRM’s ability to extract semantics, we vary the semantic encoder across three pre-trained language models of different sizes. Their model sizes and output dimensions are listed in Table~\ref{tab:encoders}.

We fix the input text to the same fields for all models: “Title”, “Price”, “Brand”, “Category”, and “Description”, and keep all other configurations fixed. Table~\ref{tab:plm_ndcg} reports NDCG@10 on the three benchmarks and shows that performance rises with the semantic encoder’s size and capacity, with DiffGRM benefiting the most.

\section{Supervision Signals for ARM and MDM}\label{app:sup_details}

We visualize the learnable supervision signals for the autoregressive model and the masked diffusion model in Figure~\ref{fig:sup_signals}. A \emph{learnable supervision signal} refers to predicting a target digit \(k\) under a specific visible–masked configuration of the other \(n{-}1\) digits.

\noindent\textbf{ARM.}
Under teacher forcing with causal masking, each digit \(k\) is supervised once with the visible set being its strict prefix. Hence one sample contributes exactly \(n\) signals. All \(n\) digit–context pairs are covered by a single sample, so the minimum number of samples is \(1\).

\noindent\textbf{MDM.}
Masked diffusion considers non-empty masked sets \(S\subseteq\{0,\ldots,n{-}1\}\). For each such \(S\), every \(k\in S\) yields one signal under the corresponding visible set. The total number of distinct target–context pairs is \(\sum_{m=1}^n \binom{n}{m} m = n\,2^{\,n-1}\). To cover all signals when each sample provides one masking pattern, the minimum number of samples is \(2^{\,n}-1\). In practice we do not enumerate all patterns, but prioritize high-value masked sets, which yields broad coverage with fewer samples.

\begin{table}[b]
\centering
\caption{Semantic encoders used in the analysis.}
\label{tab:encoders}
\begin{tabular}{lcc}
\toprule
\textbf{Encoder} & \textbf{Model size} & \textbf{Output dim.} \\
\midrule
\texttt{sentence-t5-base} & 110M & 768 \\
\texttt{bge-large-en-v1.5} & 335M & 1024 \\
\texttt{gte-large-en-v1.5} & 434M & 1024 \\
\bottomrule
\end{tabular}
\end{table}

\begin{table}[b]
\centering
\caption{NDCG@10 with different semantic encoders.}
\label{tab:plm_ndcg}
\begin{tabular}{l l c c c}
\toprule
\textbf{Model} & \textbf{semantic encoder} & \textbf{Sports} & \textbf{Beauty} & \textbf{Toys} \\
\midrule
\multirow{3}{*}{RPG} 
 & \texttt{sentence-t5-base}  & 0.0238 & 0.0429 & 0.0460 \\
 & \texttt{bge-large-en-v1.5} & 0.0248 & 0.0408 & 0.0421 \\
 & \texttt{gte-large-en-v1.5} & 0.0229 & 0.0423 & 0.0469 \\

\midrule
\multirow{3}{*}{DiffGRM} 
 & \texttt{sentence-t5-base}  & 0.0305 & 0.0502 & 0.0524 \\
 & \texttt{bge-large-en-v1.5} & 0.0327 & 0.0564 & 0.0508 \\
 & \texttt{gte-large-en-v1.5} & 0.0342 & 0.0549 & 0.0510 \\

\bottomrule
\end{tabular}
\end{table}

\begin{table}[t]
\centering
\caption{Counts with an \(n\)-layer codebook. \emph{signals}: total learnable supervision signals. \emph{samples}: minimum training samples to cover all signals when each sample provides one masking pattern.}
\label{tab:sup_combinatorics}
\setlength{\tabcolsep}{8pt}
\begin{tabular}{lcc}
\toprule
\textbf{Paradigm} & \textbf{signals} & \textbf{samples} \\
\midrule
autoregressive model & \(n\) & \(1\) \\
masked diffusion model & \(n\,2^{\,n-1}\) & \(2^{\,n}-1\) \\
\bottomrule
\end{tabular}
\end{table}

\section{ARM Sample Expansion}\label{app:arm_data_augmentation}
To compare the effect of sample expansion across paradigms, we note that under teacher forcing an autoregressive model supervises all \(n\)-digits SID of an item in a single pass, so duplicating the same training instance does not introduce new supervision signals. In contrast, a masked diffusion model supervises only the masked digits per view, and increasing coherent paths exposes more distinct supervision signals for the same training instance.

To verify this, we reimplemented an ARM-style generative recommender with the open-source GRID framework~\cite{GRID}, using the same encoder–decoder backbone as our main system and RQ-KMeans quantization. On the Toys dataset we compare two data settings: 1x uses the original training set, and 4x duplicates each training instance four times without changing targets or token order. All other hyperparameters and early stopping remain unchanged. 

The results in Table~\ref{tab:arm_dup} show negligible differences, indicating that expanding samples that carry the same supervision does not improve the ARM, and supporting that the gains of the MDM come from exposing additional supervision signals rather than adding more copies of identical supervision.

\begin{table}[t]
\centering
\caption{Effect of duplicating training instances for the ARM on Toys. 1x is the original set. 4x duplicates each instance four times with identical supervision. Expanding samples with the same supervision does not improve performance.}
\label{tab:arm_dup}
\setlength{\tabcolsep}{4pt}
\begin{tabular}{lcccc}
\toprule
\textbf{Setting} & \textbf{Recall@5} & \textbf{Recall@10} & \textbf{NDCG@5} & \textbf{NDCG@10} \\
\midrule
1x  & 0.0415 & 0.0624 & 0.0273 & 0.0341 \\
4x  & 0.0422 & 0.0627 & 0.0274 & 0.0340 \\
\bottomrule
\end{tabular}
\end{table}

\begin{table}[t]
\centering
\caption{Effect of sliding-window augmentation on training-sample count and performance.}
\label{tab:sw_aug}
\begin{tabular}{l l c c}
\toprule
\textbf{Dataset} & \textbf{Setting} & \textbf{NDCG@10} & \textbf{samples} \\
\midrule
\multirow{2}{*}{Sports}
& No sliding window  & 0.0237 & 35{,}598 \\
& Sliding window     & 0.0305 & 152{,}346 \\
\midrule
\multirow{2}{*}{Beauty}
& No sliding window  & 0.0350 & 22{,}363 \\
& Sliding window     & 0.0502 & 105{,}668 \\
\midrule
\multirow{2}{*}{Toys}
& No sliding window  & 0.0396 & 19{,}412 \\
& Sliding window     & 0.0524 & 87{,}180 \\
\bottomrule
\end{tabular}
\end{table}

\section{Sliding Window Data Augmentation}\label{app:data_augmentation}
Training samples denotes the number of sequence instances actually used for optimization after sequence splitting/augmentation. Following prior work, we adopt sliding-window augmentation, where a user interaction sequence is expanded into all possible contiguous sub-sequences \cite{DBLP:conf/www/ZhouHXGWK024,lee2025sequentialdataaugmentationgenerative}. Concretely, each subsequence forms a training instance whose target is the next item, which passes its item SIDs through our masked diffusion decoder. This exposes more item–SID correspondences and richer contexts, helping the model learn more generalizable patterns, mitigate overfitting, and remain robust under sparsity or noise. Results in Table~\ref{tab:sw_aug} show consistent gains across datasets as the training-sample count increases.

\begin{figure}[t]
  \centering
  \includegraphics[width=\columnwidth]{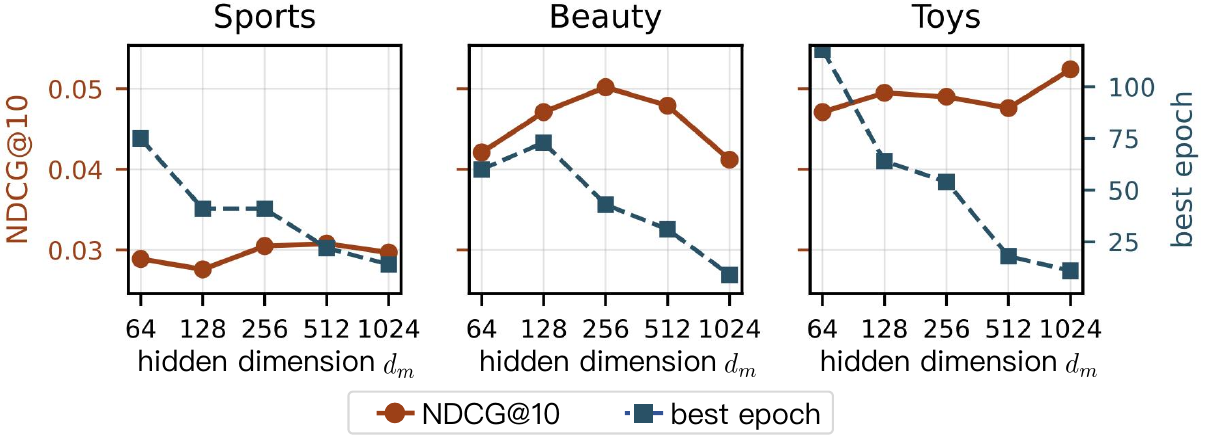}
  \caption{Analysis of DiffGRM performance (NDCG@10) and best epoch w.r.t.\ hidden dimension $d_m$.}
  \label{fig:dim}
\end{figure}

\section{Hidden Dimension Analysis}\label{sec:dim}
Figure~\ref{fig:dim} varies $d_m \in \{64,128,256,512,1024\}$ with all other settings fixed. The performance knee appears near $d_m=256$ on Sports and Beauty, whereas Toys continues to improve up to $d_m=1024$. Increasing $d_m$ also enlarges the parameter count and the associated memory and compute cost. As $d_m$ grows, the best epoch becomes smaller, indicating faster convergence. Balancing accuracy and efficiency, we set $d_m=256$ for Sports and Beauty and $d_m=1024$ for Toys, and use these settings in the main experiments.

\end{document}